\documentclass{jfm}

\def\XXint#1#2#3{{\setbox0=\hbox{$#1{#2#3}{\int}$}
     \vcenter{\hbox{$#2#3$}}\kern-.5\wd0}}

\makeatletter
\newcommand{\specialnumber}[1]{%
  \def\tagform@##1{\maketag@@@{(\ignorespaces##1\unskip\@@italiccorr#1)}}%
}
\newcommand{\specialeqref}[2]{\begingroup
  \def\tagform@##1{\maketag@@@{(\ignorespaces##1\unskip\@@italiccorr#2)}}%
  \eqref{#1}\endgroup}
\makeatother

\author[Y. A. Semenov, G.X. Wu]{ Yuriy A. Semenov, Guo Xiong Wu}
\affiliation{Author's affiliation}

\title[Impulsive impact of a submerged body]
      {Impulsive impact of a submerged body}
\author[Y. A. Semenov, Y.N. Savchenko, G.Y. Savchenko]
{Y. \ls A.\ns S\ls E\ls M\ls E\ls N\ls O\ls V\ls$^1$  \thanks{Email address for correspondence: yuriy.a.semenov@gmail.com},  Y. \ls N. \ns S\ls A\ls V\ls C\ls H\ls E\ls N\ls K\ls O\ls   $^1$,  G.\ls Y.\ns S\ls A\ls V\ls C\ls H\ls E\ls N\ls K\ls O\ls   $^1$}
\affiliation{$^1$National Academy od Science of Ukraine, Institute of Hydromechanics, Kyiv, 03057, Ukraine}

\pubyear{??}
\volume{???}
\pagerange{1--20}
\date{?? and in revised form ??}
\begin{document}
\maketitle

\begin{abstract}
An analytical solution of the impulsive impact of a cylindrical body submerged below a calm water surface is obtained by solving a free boundary problem. The shape of the cross section of the body is arbitrary. The integral hodograph method is applied to derive the complex velocity potential defined in a parameter plane. The boundary-value problem is reduced to a Fredholm integral equation of the first kind in the velocity magnitude on the free surface. The velocity field, the impulsive pressure on the body surface, and the added mass are determined in a wide range of depths of submergence for various cross-sectional shapes, such as a flat plate, a circular cylinder, and a rectangle.
\end{abstract}

\section{Introduction}
The concept of impulsive fluid/structure interaction is widely used to study the initial stage of violent water impact flows. Among the earliest work is von Karman's (1929) analytical solution of the impulsive impact of a flat plate floating on a free surface. Havelock (1949) studied an impulsively starting motion of a cylinder, with a constant velocity and a constant acceleration, respectively. He applied a linearized free surface boundary condition and investigated the full time evolution of the free surface. \cite{Tyvand1995} studied an unsteady nonlinear free-surface flow using the method of small-time series expansion taking into consideration orders high enough  to account for the leading gravitational effects on the surface elevation and predict the hydrodynamic force acting on the cylinder. The related problem of water-entry (water-exit) of a circular cylinder was studied theoretically and experimentally by \cite{Greenhow_83} and \cite{Greenhow_87}.

	The impulsive concept is used to predict wave impacts on marine and coastal structures (\cite{Cooker_1995}), ship slamming (\cite{Faltinsen2005}), the impulsive vertical motion of a body initially floating on a flat free surface (\cite{Iafr_Kor2005}), dam-break flows (\cite{Korobkin_Yilmaz}), impulsive sloshing in containers and tanks (\cite{Tyvand2012}), drops that hit a solid or liquid surface in impulsive impact (\cite{Tyvand2017}). A solution based on the impulsive concept may contain a singularity on the three-phase contact line. In such cases, the solution is used as an outer solution which has to be matched with an inner solution in the vicinity of the singularity using the method of matched asymptotic expansion (\cite{King1994, Needham2007}).

In contrast to the previous studies, we consider the impulsive motion of a body fully submerged in a liquid. The motivation for this research comes from the naval hydrodynamics of high-speed hydrofoil crafts, whose foil system may experience sudden vertical impacts caused by waves hitting the main body of the craft. The impulsive pressure on the body, the velocity on the free surface, and the associated added mass are determined in a wide range of depths of submergence and for various shapes including a flat plate, a circular cylinder, and a rectangle.

\section{Boundary-value problem} \label{sec:2}
\begin{figure}
\centering
\includegraphics[scale=0.65]{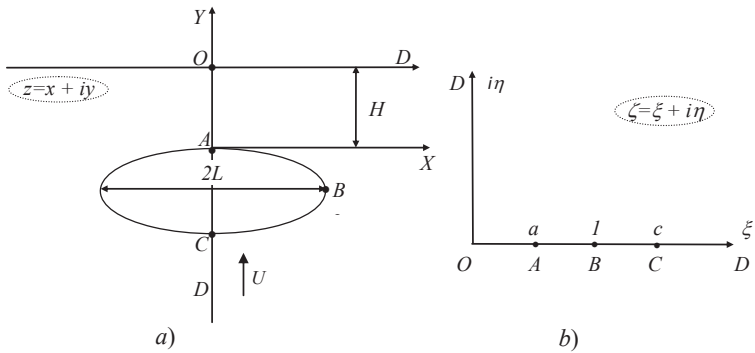}\\
\caption{($a$) Sketch of the physical plane, ($b$) the parameter, or $\zeta-$plane. }
\label{figure1}
\end{figure}

A sketch of the physical domain is shown in figure \ref{figure1}($a$). The body submerged below a calm free surface is symmetric about the $Y-$axis, and thus only a half of the flow region is considered. Before the impact, $t=0$, the body and the liquid are at rest. At time $t=0^+$ the body is suddenly set into motion with acceleration $a$ directed downward such that during infinitesimal time interval $\Delta t \rightarrow 0$, the speed of the body reaches the value $U$. The problem of a rigid body moving in a fluid body is kinematically equivalent to the problem of a fluid body moving around a fixed rigid body with acceleration $a$ at infinity. We define a noninertial Cartesian system of coordinates $XY$ attached to the body at point $A$, and an inertial system of coordinates $X^\prime Y^\prime$  in which the velocity of the liquid at infinity equals zero. The body is assumed to have an arbitrary shape, which can be defined by the slope of the body as a function of the arc length coordinate $S$, $\beta_b=\beta_b(S)$. The liquid is assumed to be ideal and incompressible. The flow starting from the rest remains irrotational at subsequent times. The gravity and surface tension are ignored.

We can introduce complex potentials $W(Z)=\Phi(X,Y)+i\Psi(X,Y)$ and $W^\prime(Z)=\Phi^\prime(X,Y)+i\Psi^\prime(X,Y)$ with $Z=X+iY$. In these systems, the velocity fields are related as follows
\begin{equation}
\label{eqv1}
\frac{dW}{dZ}=\frac{dW^\prime}{dZ} - iat,
\end{equation}
where $a$ is the acceleration,  $0 < t < \Delta t$  and $\Delta t \rightarrow 0$.  Integrating equation (\ref{eqv1}), we can find
\begin{equation}
\label{eqv2}
\specialnumber{a,b,c}
W=W^\prime - iatZ,   \quad  \frac{\partial W}{\partial t}=\frac{\partial W^\prime}{\partial t} - iaZ, \quad  \frac{\partial \Phi^\prime}{\partial t}=\frac{\partial \Phi}{\partial t} - aY.
\end{equation}

By substituting the last equation into Bernoulli's equation
\begin{equation}
\label{eqv3}
\frac{\partial \Phi^\prime}{\partial t}+\frac{p}{\rho}+ \frac{|V^\prime|^2}{2}=\frac{p_a}{\rho}
\end{equation}
integrating it over infinitesimal time interval $\Delta t\rightarrow 0$, and taking into account that the integral of the third term tends to zero, one can obtain
\begin{equation}
\label{eqv4}
P=\int_0^{\Delta t}pdt=-\rho\Phi+\rho UY,
\end{equation}
where $P$ is the impulsive pressure and  $U=a\Delta t$. Here, $|V^\prime|<\infty$ is the velocity magnitude, $p$ and $p_a$ are the hydrodynamic pressure and the pressure on the free surface, respectively.

We introduce dimensionless quantities normalized to  $U$, $L$, and $\rho$:Then,  $v=|V|/U$, $x=X/L$, $y=Y/L$, $h=H/L$, $\phi(s)=\Phi(S)/(LU)$. The vertical impulse force $F_y$ is obtained by integrating the impulse pressure over the body surface,
\begin{equation}
\label{eqv5}
F_y=-2\rho L^2 U\int_{s_A}^{s_C}\phi(s)\cos(n,y) ds-2\rho UA=\rho mL^2 U,		
\end{equation}
where $s$ is the arc length coordinate along the body surface, $s_A$  and $s_C$  are the arc lengths of points $A$  and $C$, $m$ is the coefficient of the added mass and $A$ is the cross-sectional area . The factor $"2"$ accounts for the force acting on the part of the body symmetric about the $y-$axis. Due to the symmetry of the body about the $y$-axis, the horizontal impulse force equals zero. 

It is desired to determine the velocity potential $\phi(s)$ immediately after the impact.

\section{Conformal mapping} \label{sec:3}
A direct finding of the complex potential $w(z)$ is a challenge; therefore, we introduce an auxiliary parameter plane, or $\zeta-$plane as suggested by \cite{Michell_1890} and \cite{Joukovskii_1890}. We formulate boundary-value problems for the complex velocity function, $dw/dz$, and for the derivative of the complex potential, $dw/d\zeta$, both defined in the  $\zeta-$plane. Then the derivative of the mapping function is obtained as $dz/d\zeta  = (dw/d\zeta )/(dw/dz)$, and its integration provides the mapping function $z = z(\zeta)$ relating the coordinates in the parameter and physical planes.

We choose the first quadrant of the $\zeta-$plane in figure \ref{figure1}$b$ as the region corresponding to the fluid region in the physical plane (figure \ref{figure1}$a$). The conformal mapping theorem allows us to arbitrarily choose the locations of three points, which are points $O$ at the origin ($\zeta=0$), $D$ ($D^\prime$) at infinity, and $B$ at $\zeta= 1$ (see figure \ref{figure1}$b$). The position of points $A$ ($\zeta=a$) and $C$ ($\zeta=c$) has to be determined from the solution of the problem and physical considerations.

\subsection{Expressions for the complex velocity and the derivative of the complex potential} \label{subsec:31}
	The body is considered to be fixed; therefore, the velocity direction and the slope of the body coincide. Besides, at this stage we assume that the velocity magnitude on the free surface is a known function of the parameter variable, $v(\eta)$. Then the boundary-value problem for the complex velocity in the first quadrant of the parameter plane can be written as follows
\begin{equation}
\label{eqv7}
\chi (\xi ) = \arg \left(\frac{dw}{dz} \right) = \left\{{\begin{array}{l}
- \beta_b(a) + \beta_0,\quad 0 \leq \xi \leq a, \\
- \beta_b(\xi),\qquad\quad\;  a \leq \xi \leq c, \\
- \beta_b(c) - \beta_0,\quad c \leq \xi < \infty.
\end{array}} \right.
\end{equation}
\begin{equation}
\label{eqv8} v(\eta ) = \left| {\frac{dw}{dz}} \right|_{\zeta=i\eta}, \qquad 0\leq\eta<\infty.
\end{equation}
where $\beta_0=\pi/2$,  $\beta_b(a)=\pi$ and $\beta_b(c)=0$. Equation (\ref{eqv7}) satisfies the conditions:  $\chi(\xi)=-\pi/2$ along the intervals $OA$ and $CD$ on the symmetry line, and $\chi(\xi)=-\beta_b(\xi)$ along the body. The argument of the complex velocity exhibits jumps $\Delta=-\pi/2$ at points $A$ and $C$ when we move along the boundary in the physical plane from point $O$ to point $D$. This boundary-value problem can be solved by applying the following integral formula (Semenov \& Yoon, 2009):
\begin{equation}
\label{eqv9}
\frac{dw}{dz} = v_{\infty}\exp \left[ {\frac{1}{\pi }\int\limits_0^\infty {\frac{d\chi }{d{\xi }}\ln \left(
{\frac{\zeta + {\xi }}{\zeta - {\xi }}} \right)d{\xi } - \frac{i}{\pi }\int\limits_0^\infty {\frac{d\ln v}{d{\eta }}\ln
\left( {\frac{\zeta - i{\eta }}{\zeta + i{\eta }}} \right)d{\eta } + i\chi_{\infty}}}} \right],
\end{equation}
where $v_\infty=v(\eta)_{\eta\rightarrow \infty}$ and  $\chi_\infty=\chi(\xi)_{\xi \rightarrow \infty}$. Substituting equations (\ref{eqv7}) and (\ref{eqv8}) into (\ref{eqv9}) and evaluating the first integral over the step change at points $\zeta=a$ and $\zeta=c$, we obtain
\begin{eqnarray}
\label{eqv10}
\frac{dw}{dz} &=& v_\infty\left(\frac{\zeta - a}{\zeta + a}\right)^{\frac{1}{2}}\left(\frac{\zeta - c}{\zeta + c}\right)^{\frac{1}{2}} \\ \nonumber
&\times& \exp\left[\frac{1}{\pi}\int\limits_a^c{\frac{d\beta_b}{d{ \xi}}\ln \left(\frac{\zeta - \xi}{\zeta + \xi}\right) d{\xi}}
- \frac{i}{\pi}\int\limits_0^\infty {\frac{d\ln v}{d{\eta }}\ln \left( {\frac{\zeta - i{\eta }}{\zeta + i{\eta }}} \right)d{\eta}} - i\beta_0 \right].
\end{eqnarray}
It can be easily verified that for $\zeta=\xi$ the argument of the right-hand side of (\ref{eqv10}) is the function $-\beta_b(\xi)$, while for $\zeta=i\eta$ the modulus of (\ref{eqv10}) is the function $v(\eta)$,  i.e. the boundary conditions (\ref{eqv7}) and (\ref{eqv8}) are satisfied. It can also be seen that the complex velocity function has zeros of order $1/2$, which  correspond to the flow around a corner of angle  $\pi/2$ at points $A$ and $C$.

In order to derive the derivative of the complex potential, we have to analyse its behaviour. Before the impact, the free surface is flat, and it coincides with the $x-$axis (see figure \ref{figure1}$a$). The pressure along the free surface is constant. Then, as follows from the Euler equation, the velocity generated by the impact is perpendicular to the free surface where the pressure is constant, or the velocity is directed in the $y-$direction. Therefore, the $x-$component of the velocity is zero, and $dw=(dw/dz) dz=(v_x-iv_y )dx=-ivdx$.  Thus, the free surface corresponds to the interval $(-\infty,0)$ on the imaginary axis $\psi$ in the $w-$plane. $\Im(w)=0$ along the line $OABCD$ due to the impermeability condition on the body $ABC$ and the intervals $OA$ and $CD$ of the symmetry line. At the same time, $\Re(w)$ changes from zero at point $O$ to $-\infty$ at infinity $D$. Thus, the flow region in the physical plane corresponds to the third quadrant in the $w-$plane. They are related as  $w=-K\zeta$ where $K$ is a positive real number. Then one can  obtain
\begin{equation}
\label{eqv11}
\frac{dw}{d\zeta}=-K.
\end{equation}
The simple form of the complex potential, $w(\zeta)=-K\zeta+w_O$, allows one to exclude the parameter $\zeta$ and obtain the solution in the Kirchhoff's form, for which the complex velocity is the function of the complex potential. Here, $w_O$ is the complex potential at point $O$.

The derivative of the mapping function is obtained by dividing (\ref{eqv11}) by (\ref{eqv10})
\begin{eqnarray}
\label{eqv12}
\frac{dz}{d\zeta} &=& -K\left(\frac{\zeta + a}{\zeta - a}\right)^{\frac{1}{2}}\left(\frac{\zeta + c}{\zeta - c}\right)^{\frac{1}{2}} \\ \nonumber
&\times& \exp\left[\frac{1}{\pi}\int\limits_a^c{\frac{-d\beta_b}{d{\xi}}\ln \left(\frac{\zeta - \xi}{\zeta + \xi}\right) d{\xi}}
+ \frac{i}{\pi}\int\limits_0^\infty {\frac{d\ln v}{d{\eta }}\ln \left( {\frac{\zeta - i{\eta }}{\zeta + i{\eta }}} \right)d{\eta}} + i\beta_0 \right].
\end{eqnarray}
The integration of this equation yields the mapping function $z = z(\zeta)$ relating the parameter and the physical planes. Equations (\ref{eqv10}) and (\ref{eqv11}) include the parameters  $a$, $c$, $K$ and the functions  $v(\eta)$ and $\beta_b(\xi)$, all to be determined from physical considerations and the kinematic boundary condition on the free surface and on the solid boundary $OABCD$.
\subsection{Body boundary conditions} \label{subsec:32}
The arc lengths between points $A$ and $B$, $s_{AB}$, and between points $B$ and $C$, $s_{BC}$, and the depth of submergence are determined as follows
\begin{equation}
\label{eqv15}
\int_a^1\left|\frac{dz}{d\zeta} \right|_{\zeta=\xi}=s_{AB},         \qquad  \int_1^c\left|\frac{dz}{d\zeta} \right|_{\zeta=\xi}=s_{BC}, \qquad \int_0^a\left|\frac{dz}{d\zeta} \right|_{\zeta=\xi}=h.
\end{equation}
The unknown function $\beta_b (\xi)$ is determined from the following integro-differential equation
\begin{equation}
\label{eqv16}
\frac{d\beta_b}{d\xi} = \frac{d\beta_b}{ds} \left|\frac{dz}{d\zeta} \right|_{\zeta=\xi},
\end{equation}
where $\beta_b(s)$ is the given function. Equation (\ref{eqv16}) is solved by iteration using $d\beta_b/d\xi$  in (\ref{eqv12}) known at the previous iteration.
\subsection{Free surface boundary conditions} \label{subsec:33}
An impulsive impact is characterized by infinitesimally small time interval $\Delta t\rightarrow 0$ such that the position of the free surface does not change during the impact. From the Euler equations it follows that the velocity generated by the impact is perpendicular to the free surface ($p=p_a$)
\begin{equation}
\label{eqv17}
\arg \left(\left. \frac{dw}{dz}\right|_{\zeta=i\eta} \right)=-\beta_0, \qquad   0 \le \eta \le \infty.
\end{equation}
Taking the argument of the complex velocity from (\ref{eqv10}), we obtain the following integral equation in the function $d\ln v/d\eta$
\begin{equation}
\label{eqv18}
\int\limits_0^\infty \frac{d\ln v}{d\eta }\ln \left| \frac{\eta^\prime-\eta}{\eta^\prime+\eta} \right|d\eta^\prime+\tan^{-1}\left(\frac{\eta}{a}\right) + \tan^{-1}\left(\frac{\eta}{c}\right) + \frac{2}{\pi}\int_a^c \frac{d\beta_b}{d\xi} \tan^{-1}\left(\frac{\eta}{\xi}\right)d\xi =0
\end{equation}

Equation (\ref{eqv18}) is a Fredholm integral equation of the first kind with a logarithmic kernel. Its solution takes the form (see Appendix)
\begin{equation}
\label{eqv19}
v(\eta)=\sqrt{\eta^2+a^2}\sqrt{\eta^2+c^2}\exp\left(\frac{1}{\pi}\int_a^c\frac{d\beta_b}{d\xi}\ln (\eta^2+\xi^2)d\xi\right).
\end{equation}
The system of equations (\ref{eqv15}), (\ref{eqv16}) and (\ref{eqv18}) forms a closed system of equations in the parameters $a$, $c$, $K$  and functions $\beta_b(\xi)$ and $\ln v(\eta)$.

\section{Results} \label{sec:4}
\begin{figure}[]
\centering
\includegraphics[scale=0.4]{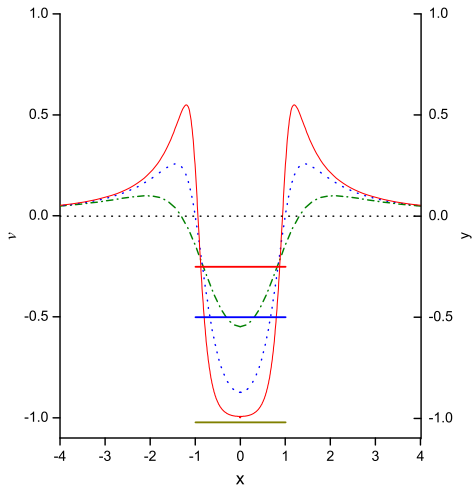}
\vspace*{-20mm}
\caption{Velocity magnitude on the free surface (left axis) for various depths of submergence (right axis): h=0.25 (red, solid line); 0.5 (blue, dashed line); 1.0 (olive, dot-dashed line)}
\label{figure2}
\end{figure}
The results presented below are shown in the system of coordinates related to the liquid at infinity. The velocity distribution on the free surface generated by the impulsive impact of a flat plate is shown in figure \ref{figure2} for various depths of submergence. The position of the plate (right axis) is shown as a thick horizontal line. The colours of the plate and the associated velocity distribution are the same. After the impact, the plate gets a velocity equal to $-1$. At a relatively small depth of submergence, $h<0.25$, the velocity of the liquid above the plate, $|x|<1$,  is close to $-1$, i.e. the plate entrains the liquid, and they move almost together. Outside of the plate, $|x|>1$, the plate displaces the liquid, and it moves upward providing the balance of the liquid coming into and out of the flow region. The larger the depth of submergence of the plate, the smaller the response of the free surface.

\begin{figure}
\centering
\includegraphics[scale=0.35]{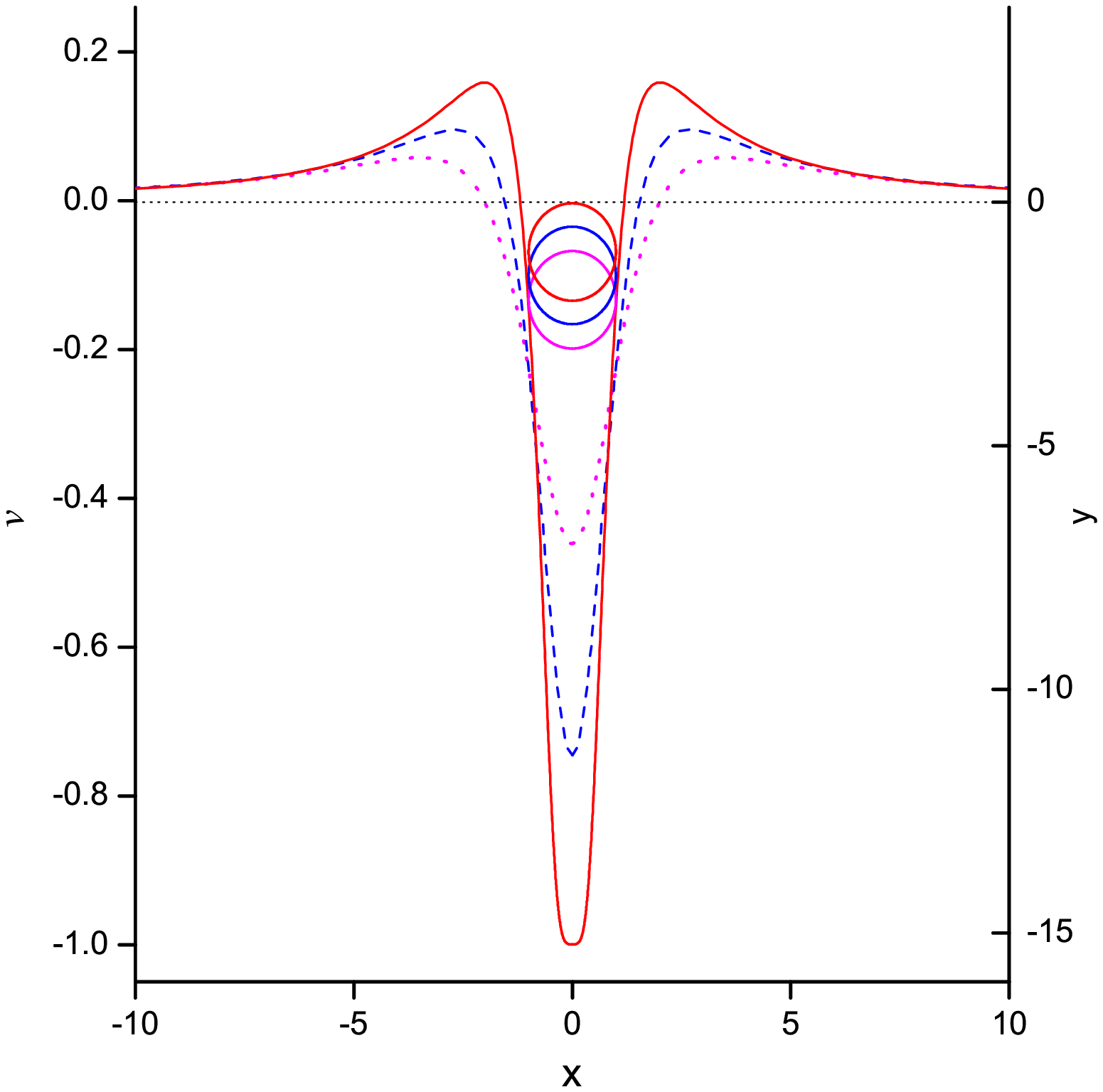}
\hspace*{-10mm}
\includegraphics[scale=0.35]{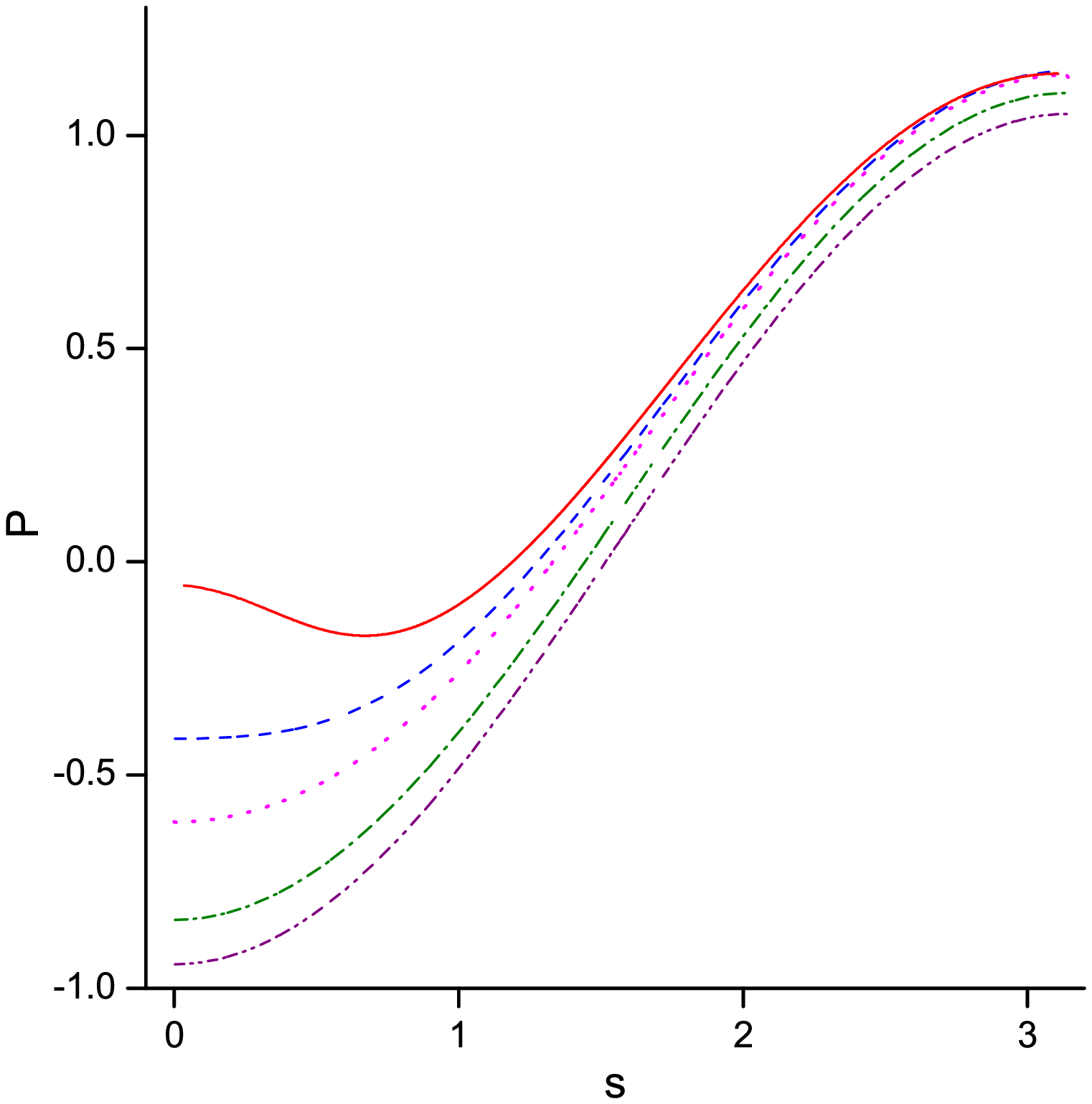}\
\vspace*{-18mm}
\caption{($a$) velocity along the free surface (left axis) ($b$) impulsive pressure along the cylinder for various depths of submergence: $h=0.02$ (red, solid line), $0.1$ (blue, dashed line), $1.0$ (magenta, dotted line),  $3.0$ (green, dot dashed) and $10.0$ (purple, dash-dot-dot)}
\label{figure3}
\end{figure}
The velocity distribution along the free surface and the impulsive pressure versus the arc length coordinate $s$ are shown for a circular cylinder in figures \ref{figure3}$a$ and \ref{figure3}$b$, respectively. The coordinates $s=0$ and $s=pi$ correspond to the top (point $A$) and the bottom (point $C$) of the cylinder. The bottom pressure is positive, and it depends only weakly on the depth of submergence. At the top of the cylinder, the pressure is negative, but it increases as the cylinder approaches the free surface.

\begin{figure}
\centering
\includegraphics[scale=0.4]{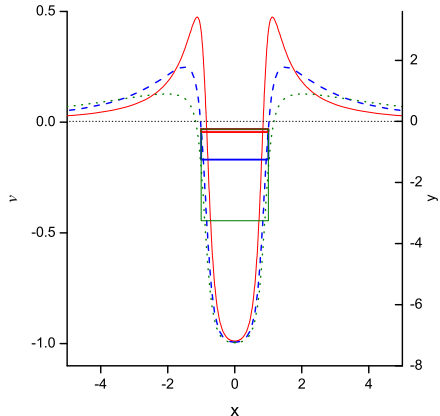}
\vspace*{-23mm}
\caption{Velocity along the free surface generated by the impact of a rectangle submerged at depth $h=0.25$ (left axis) and the position of the rectangle (right axis): side length $b=0.1$ (red, solid line), $m=2.272$; $b=1.0$ (blue, dashed line) , $m= 2.786$; $b=3$ (magenta, dotted line), $m= 3.241$.}
\label{figure4}
\end{figure}

The response of the free surface caused by the impulsive impact of a submerged rectangle is shown in figure \ref{figure4} for depth $h=0.25$ and various values of the side length $b$ of the rectangle. For a small length, $b=0.1$, the velocity distribution is close to that shown in figure \ref{figure2} for a plate at depth $h=0.25$. For larger lengths $b$, the velocity decreases, while the length of the free surface in $x-$direction affected by the impact increases. Such behaviour provides a  balance between the liquid moving into and out of the flow region.

	The added mass coefficients are shown in table 1 for various shapes of the body. For a flat plate, $m\rightarrow\pi/2$ as $h\rightarrow0$,  which agrees with the added mass for a flat plate floating on a free surface (von Karman impact solution). For a large depth of submergence, $h=50$, the effect of the free surface becomes negligible, and the coefficient of the added mass approaches the value corresponding to the added mass in an unbounded fluid domain.
\begin{table}
  \begin{center}
\def~{\hphantom{0}}
  \begin{tabular}{lcccccccccc}
      $h$    & $0$     &  $0.02$ & $0.05$ & $0.1$ & $0.3$ & $0.5$ & $1$   & $5$   & $50$  & $\infty^\ast$ \\[3pt]
   $Plate$   & 1.571   & ~~1.624~& 1.735  & 1.876 & 2.265 & 2.516 & 2.835 & 3.108 & 3.137 & $\pi$    \\
   $Circle$  & -       & ~~2.090 & 2.111  & 2.162 & 2.370 & 2.531 & 2.777 & 3.104 & 3.141 & $\pi$    \\
   $Squire$  & -       & ~~2.993 & 3.011  & 3.048 & 3.292 & 3.552 & 4.024 & 4.667 & 4.754 & 4.754  \\
  \end{tabular}
  \caption{Added mass coefficient for various body shapes and depths of submergence. $^\ast$ unbounded fluid domain}.
  \label{tab:kd}
  \end{center}
\end{table}

\section{Upward impulsive impact} \label{sec:5}
Here, we analyse how the direction of an impact on a submerged body affects the velocity field. In the system of coordinates attached to the body, a change of the impact direction results in a reversal of the velocity direction of the liquid  at infinity and on the whole solid boundary including the body and the symmetry line. Equation (\ref{eqv7}) will keep its form if we set values  $\beta_0=-\pi/2$,  $\beta_b (a)=0$  and $\beta_b(c)=-\pi$, or subtract $\pi$ from $\chi(\xi)$ in (\ref{eqv7}). In this case,  expression for the complex velocity (\ref{eqv10}) keeps its form.   The derivative of the complex potential is obtained from (\ref{eqv11}) if we change the sign of the expression. Thus, all the equations of the problem keep their form. Therefore, the velocity magnitude on the free surface (\ref{eqv19}) remains the same for both upward and downward impact directions.

\begin{figure}
\centering
\includegraphics[scale=0.4]{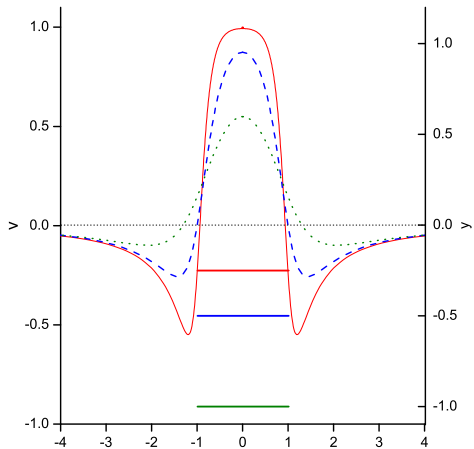}
\vspace*{-23mm}
\caption{Same as in figure 2 but opposite (upward) direction of the impact.}
\label{figure5}
\end{figure}

In the system of coordinates related to the liquid, a change of the impact direction results in a reversal of the velocity direction on the free surface. Figure \ref{figure5} illustrates the velocity distribution on the free surface corresponding to an upward impact of the plate for the same depths of submergence as shown in figure \ref{figure2} for a downward impact.

\section{Conclusions} \label{sec:6}
An analytical solution of the impulsive impact of a cylindrical body below an undisturbed free surface is found using the integral hodograph method. The cross-sectional shape of the body may be a polygon or an arbitrary smooth shape. The results are shown for a flat plate, a circular cylinder, and a rectangle. The equation for the complex velocity includes the velocity magnitude on the free surface, whose analytic form has been determined.

The associated added masses are found as a function of the depth of submergence. As the depth of submergence tends to infinity, the added mass tends to the value corresponding to the added mass in an unbounded fluid domain.

It is shown that upward and downward impacts generate an identical magnitude of the velocity on the free surface and identical added mass coefficients. However, the velocity direction is opposite. The obtained solution can be considered as a first-order solution in solving the problem using the method of small-time series.

\section*{Declaration of interests}
The authors report no conflict of interests.

\appendix
\section{}
Let's consider a polygon inscribed in a given smooth boundary of the body. Let the $N_p$ be the number of the sides of length $l_i$ the slope to which is $\beta_{pi}$. Let $\xi_i$,  $i=1,2, \ldots,N_p$, be the points in the parameter region corresponding to the vertexes of the polygon. Then the function $\beta_b(\xi)$ can be written explicitly
\begin{equation}
\label{eqv20}
\beta_b(\xi)=\beta_{bi}, \qquad  \xi_{i-1} < \xi \le \xi_i, \quad i=1, \ldots N_p,
\end{equation}
where $\xi_0=a$,  $\xi_{N_p}=c$,  $\beta_{p1}=\pi$ and $\beta_{Np}=0$.

By substituting (\ref{eqv20}) into (\ref{eqv10}) and evaluating the integral over the step change in the function $\beta_b(\xi)$ at points $\xi=\xi_i$, we obtain the complex velocity for the polygon-shaped body
\begin{eqnarray}
\label{eqv21}
\frac{dw}{dz} &=& \left(\frac{\zeta - a}{\zeta + a}\right)^{\frac{1}{2}}\left(\frac{\zeta - c}{\zeta + c}\right)^{\frac{1}{2}} \prod_{i=1}^{N_p}\left(\frac{\zeta - \xi_i}{\zeta + \xi_i}\right)^{\frac{\Delta\beta_{pi}}{\pi}} \\ \nonumber
&\times& \exp\left[- \frac{i}{\pi}\int\limits_0^\infty {\frac{d\ln v}{d{\eta }}\ln \left( {\frac{\zeta - i{\eta }}{\zeta + i{\eta }}} \right)d{\eta}} - i\beta_0 \right].
\end{eqnarray}
The free-surface boundary condition (\ref{eqv17}) with the complex velocity (\ref{eqv21}) leads to the integral equation
\begin{equation}
\label{eqv27}
\int\limits_0^\infty \frac{d\ln v}{d\eta }\ln \left| \frac{\eta^\prime+\eta}{\eta^\prime-\eta} \right|d\eta^\prime=f(\eta),
\end{equation}
where
\[f(\eta)=\tan^{-1}\left(\frac{\eta}{a}\right) + \tan^{-1}\left(\frac{\eta}{c}\right) + \frac{2}{\pi}\sum_{i=1}^{N_b}\Delta \beta_{bi} \tan^{-1}\left(\frac{\eta}{\xi_i}\right).
\]
By applying the following transformations (\cite{Pol_Man})
\begin{equation}
\label{eqv28}
\frac{d\ln v}{d\eta}=-\frac{2}{\pi^2}\frac{d}{d\eta}\int_\eta^\infty\frac{F(u)du}{\sqrt{u^2-\eta^2}}, \quad F(u)=\frac{d}{du}\int_0^u\frac{pf(p)}{\sqrt{u^2-p^2}}dp,
\end{equation}
the solution of the integral equation (\ref{eqv27}) is obtained
\begin{equation}
\label{eqv28}
v(\eta)=\sqrt{\eta^2+a^2}\sqrt{\eta^2+c^2}\prod_{i=1}^{N_p}\left(\eta^2 + {\xi_i}^2\right)^{\frac{\Delta\beta_{bi}}{\pi}}.
\end{equation}
By taking the limit of (\ref{eqv28}) for $N_b \rightarrow\infty$ and using $\Delta\beta_{bi}=\left(\frac{d\beta_b}{d\xi}\right)_i{\Delta\xi}_i$, we obtain equation (\ref{eqv19}).


\begin{thebibliography}{}
\bibitem[Cooker(1995)]{Cooker_1995}
{\sc Cooker, M. J., Peregrine, D.H.} 1995 Pressure-impulse theory for liquid impact problems, {\it J. Fluid Mech.} {\bf 297}, pp. 193 -- 214.

\bibitem[Faltinsen(2005)]{Faltinsen2005}
{\sc Faltinsen, O. M.} 2005 {\it Hydrodynamics of High-speed Marine Vehicles.} Cambridge University Press, 454 pp.

\bibitem[Greenhow and Lin (1983)]{Greenhow_83}
{\sc Greenhow, M.} 1983 Nonlinear free surface effects: experiments and theory. {\it Rep. 83-19. MIT, Dept. of Ocean Engineering.}

\bibitem[Greenhow(1987)]{Greenhow_87}
{\sc Greenhow, M. and Yanbao, L.} 1987 Added masses for circular cylinders near or penetrating fluid boundaries-review, extension and application to water-entry, -exit and slamming. {\it Ocean Engineering}, {\bf 14} (4). pp. 325 -- 348.

\bibitem[Havelock(1949)]{HAV_1949a}
{\sc Havelock, T. H.} 1949{\it{a}} The wave resistance of a cylinder started from rest. {\it Q. J. Mech. Appl. Maths}, {\bf 2}, pp. 325 -- 334.


\bibitem[Hjelmervik and Tyvand(2017)]{Tyvand2017}
{\sc Hjelmervik, K.B., Tyvand, P.A.} 2017 Incompressible impulsive wall impact of liquid cylinders. {\it J. Eng Math} {\bf 103}, 159 -- 171. https://doi.org/10.1007/s10665-016-9866-6

\bibitem[Joukovskii(1890)]{Joukovskii_1890}
{\sc Joukovskii, N. E. } 1890 Modification of Kirchhoff's method for determination of a fluid motion in two directions at a fixed velocity given on the unknown streamline. {\it Math. Sbornik.} {\bf 15} (1), pp. 121 -- 278.

\bibitem[von Karman(1929)]{Karman}
\sc{ von Karman, T.} 1929 The impact of seaplane floats during landing. {\it Washington, DC:NACA Tech. Note} {\bf 321.}

\bibitem[King and Needham (1994)]{King1994}
\sc{ King, A., Needham, D.} 1994 The initial development of a jet caused by fluid, body and free-surface interaction. Part 1. A  uniformly accelerating plate. {\it J. of Fluid Mech.} {\bf 268} pp. 89 -- 101.

\bibitem[Michell(1890)]{Michell_1890}
{\sc Michell, J. H.} 1890 On the theory of free stream lines, {\it Phil. Trans. R. Soc. Lond. A} {\bf 181}, 389-431.

\bibitem[Needham et al. (2007)]{Needham2007}
{\sc Needham, D., Billingham, J., King, A.} 2007 The initial development of a jet caused by fluid, body and free-surface interaction. Part 2. An impulsively moved plate. {\it J. Fluid Mech.} {\bf 578} pp. 67 -- 84.

\bibitem[Korobkin and Yilmaz (2009)] {Korobkin_Yilmaz}
{\sc Korobkin, A., Yilmaz, O.} 2009 The initial stage of dam-break flow. {\it J Eng Math} {\bf 63}, 293-308.

\bibitem[Iafrati and Korobkin (2005)] {Iafr_Kor2005}
{\sc Iafrati, A., Korobkin, A.A.} 2005 Starting flow generated by the impulsive start of a floating wedge. {\it J Eng Math} {\bf 51}, 99 -- 125.

\bibitem[Korobkin and Scolan (2006)] {Korobkin_Scolan}
{\sc Korobkin, A. A.  and Scolan, Y.-M.} 2006 Three-dimensional theory of water impact. Part 2. Linearized Wagner problem. {\it J. of Fluid Mech}, {\bf 549}, pp. 343 - 373.

\bibitem[Newman(1977)] {Newman}
{\sc Newman, J.N.} 1977 {\it Marine Hydrodynamics.}  The MIT Press, Cambridge, Massuchesetts. 389 p.

\bibitem[Oliver(2007)] {Oliver07}
{\sc Oliver J.M.} 2007 Second-order Wagner theory for two-dimensional water-entry problems at small deadrise angles. {\it J. Fluid Mech.} {\bf 572}, pp. 59 - 85.

\bibitem[Polyanin and Manzhirov (1998)]{Pol_Man}
{\sc Polyanin, A. D. and Manzhirov, A. V.} 1998 {\it Hand book of Integral Equations}. CRC Press Boca Raton London New York Washington, D.C.

\bibitem[Semenov \& Yoon (2009)]{Sem_Yoon} \textsc{Semenov, Y. A. \& Yoon, B-S.} 2009 Onset of flow separation for the oblique water impact of a wedge. \emph{Phys. of Fluids} \textbf{21}, 112103.

\bibitem[Tyvand and Miloh(1995)] {Tyvand1995}
{\sc Tyvand, P. A. and Miloh, T.} 1995 Free-surface flow due to impulsive motion of a submerged circular cylinder. {\it J. Fluid Mech.} {\bf 286}, pp. 67-101.
\bibitem[Tyvand and Miloh(2012)] {Tyvand2012}
{\sc Tyvand, P.A. and Miloh, T.} 2012 Incompressible impulsive sloshing {\it J. Fluid Mech.} {\bf 708}, pp. 279-302.


\end{thebibliography}
\end{document}